\documentclass[apjl]{emulateapj}                     

\newenvironment{inlinefigure}{
\def\@captype{figure}
\noindent\begin{minipage}{0.999\linewidth}\begin{center}}
{\end{center}\end{minipage}}

\newcommand{\ergscm}{ergs~s$^{-1}$~cm$^{-2}$}
\newcommand{\HST}{{\sl HST}}

\newcommand{\zband}{z_{850}}
\newcommand{\rp}{r_{\rm P}}

\slugcomment{{\it Astrophysical Journal Letters}, in press}

\shorttitle{AGN Morphology and Environment at $z\sim 0.4$--1.3}
\shortauthors{Grogin et al.}

\begin{document}
\setlength{\topmargin}{-0.1in}
\setlength{\textheight}{9.6in}
\title{AGN Host Galaxies at $z\sim$\,0.4--1.3\,: Bulge-Dominated and 
Lacking Merger--AGN Connection }

\author{N.~A.~Grogin\altaffilmark{1}, C.~J.~Conselice\altaffilmark{2},
E.~Chatzichristou\altaffilmark{3}, D.~M.~Alexander\altaffilmark{4},
F.~E.~Bauer\altaffilmark{5}, A.~E.~Hornschemeier\altaffilmark{1},
S.~Jogee\altaffilmark{6}, A.~M.~Koekemoer\altaffilmark{7},
V.~G.~Laidler\altaffilmark{7}, M.~Livio\altaffilmark{7},
R.~A.~Lucas\altaffilmark{7}, M.~Paolillo\altaffilmark{8},
S.~Ravindranath\altaffilmark{7}, E.~J.~Schreier\altaffilmark{9},
B.~D.~Simmons\altaffilmark{3}, and C.~M.~Urry\altaffilmark{3}}

\altaffiltext{1}{Department of Physics and Astronomy, Johns Hopkins
University, Charles and 34th Street, Baltimore, MD 21218.}
\altaffiltext{2}{Department of Astronomy, California Institute of
Technology, 1201 East California Boulevard, Pasadena, CA 91125.}
\altaffiltext{3}{Department of Astronomy, Yale University, P.O.~Box
208101, New Haven, CT 06520-8101.}
\altaffiltext{4}{Institute of Astronomy, Madingley Road, Cambridge 
CB3 0HA, UK.}
\altaffiltext{5}{Department of Astronomy, Columbia University, 550 W. 120th
Street, New York, NY 10027.}
\altaffiltext{6}{Department of Astronomy, University of Texas,
1 University Station, C1400, Austin, TX 78712-0259.}
\altaffiltext{7}{Space Telescope Science Institute, 3700 San Martin
Drive, Baltimore, MD 21218.}
\altaffiltext{8}{Dipartimento di Scienze Fisiche, Universit\'a degli Studi Federico II, 
C.~U.~Monte S.~Angelo, via Cintia, I-80126 Naples, Italy.}
\altaffiltext{9}{Associated Universities, Inc., 1400 16th Street, NW,
Washington, DC 20036.}

\clearpage
\begin{abstract}
  We investigate morphological structure parameters and local
  environments of distant moderate-luminosity active galactic nucleus
  (AGN) host galaxies in the overlap between the \HST/ACS observations
  of the Great Observatories Origins Deep Survey (GOODS) and the two
  Chandra Deep Fields.  We compute near-neighbor counts and $BViz$
  asymmetry ($A$) and concentration ($C$) indices for
  $\approx\!35,500$ GOODS/ACS galaxies complete to
  $\zband\approx26.6$, including the resolved hosts of $322$
  X-ray--selected AGNs.  Distributions of (1) $\zband$ asymmetry for
  130 $\zband<23$ AGN hosts and (2) near-neighbor counts for 173
  $\zband<24$ AGN hosts are both consistent with non-AGN control
  samples.  This implies no close connection between recent galaxy
  mergers and moderate-luminosity AGN activity out to appreciable
  look-back times ($z\lesssim 1.3$), approaching the epoch of peak AGN
  activity in the universe.  The distribution of $\zband$ $C$ for the
  AGN hosts is offset by $\Delta C\approx+0.5$ compared to the
  non-AGN, a 6.4 $\sigma$ discrepancy much larger than can be
  explained by the possible influence of unresolved emission from the
  AGN or a circumnuclear starburst.  The local universe association
  between AGN and bulge-dominated galaxies thus persists to
  substantial look-back time.  We discuss implications in the context
  of the low-redshift supermassive central black hole mass correlation
  with host galaxy properties, including concentration.
\end{abstract}


\clearpage
\section{Introduction}
The connection between active galactic nuclei (AGNs) and their host
galaxies, and the evolution in that relationship over cosmic time,
have attracted great interest in recent years.  This includes the
discoveries that most nearby massive galaxies harbor central
supermassive black holes (SMBHs; Magorrian et al.~1998), that AGNs in
the local universe ($z\lesssim1$) reside predominantly in massive,
bulge-dominated host galaxies (Kauffmann et al.~2003), and that a
tight correlation exists locally between SMBH mass and host galaxy
properties such as bulge velocity dispersion and light-profile
concentration (Ferrarese \& Merritt~2000; Gebhardt et al.~2000; Graham
et al.~2001).  The 1--2 Ms Chandra Deep Fields (CDF-N and CDF-S;
Brandt et al.~2001; Giacconi et al.~2002), which have now resolved
much of the cosmic X-ray background into moderate-luminosity AGNs at
$z \sim 1$ (Alexander et al.~2003; Barger et al.~2003; Szokoly et
al.~2004), provide a unique AGN sample to probe these locally observed
SMBH-host relationships out to epochs nearing the peaks of star
formation and AGN activity in the universe.  This is one of the aims
of the Great Observatories Origins Deep Survey (GOODS; Giavalisco et
al.~2004), which has obtained deep multicolor \textsl{Hubble Space
Telescope} (\HST) Advanced Camera for Surveys (ACS) image mosaics
across the most sensitive regions of the CDF areas.  In this Letter we
report on the local environments and rest-frame optical morphologies
of AGN host galaxies in the GOODS fields as well as the implications for
SMBH-galaxy coevolution and the merger-AGN connection.

The largest pre-GOODS investigation of \HST-imaged CDF sources was
Koekemoer et al.~(2002), with 41 CDF-S 1~Ms sources in three
moderately deep Wide Field Planetary Camera 2 (WFPC2) pointings.  Grogin et al.~(2003, hereafter G03)
studied the \HST\ morphologies and local environments of these
faintest X-ray sources as compared with the X-ray undetected
population.  The AGNs were preferentially located in galaxies with
highly concentrated light profiles, but the AGN hosts could not be
differentiated from the non-AGN based on light-profile asymmetry or
frequency of near-neighbors.  The G03 conclusions were as follows: (1) distant
moderate-luminosity AGNs did not show a connection between recent
($\lesssim\!1$~Gyr) galaxy merger/interaction and AGN activity; and (2)
the $z\sim1$ galaxy population already showed evidence of the SMBH-bulge
correlation.  Now that deeper and much larger area GOODS imaging is
available across both CDFs, we verify these results with much larger
samples of both CDF AGNs and quiescent galaxies.  We also place new
constraints on the \textit{evolution} in merger-AGN connection and
SMBH-bulge correlation with the extensive redshift information
now accumulated in these fields.  We adopt a cosmology with
$H_0=70$~km s${}^{-1}$ Mpc${}^{-1}$, $\Omega_m=0.3$, and
$\Omega_\Lambda \equiv 1 - \Omega_m = 0.7$.  Magnitudes are given in
the AB system.

\section{Observations and Sample Selection}\label{obsampsec}
Our analyses employ the \HST/ACS image mosaics in F606W ($V$), F775W ($i$), 
and F850LP ($z_{850}$) from the first three epochs of GOODS \HST\ observations
in both the northern (``GOODS-N'') and southern (``GOODS-S'') fields
(Giavalisco et al.~2004).  We use the $\zband$-detected source
catalog detailed in Ravindranath et al.~(2004), trimmed
by applying minimum thresholds in compactness ($>\!4$ pixels)
and signal-to-noise ratio ($>\!5$) optimized for removing spurious
detections.  The remaining 16,632 GOODS-S sources and 18,878
GOODS-N sources are complete to $\zband\approx26.6$ and form the basis of our
environmental and structural parameter analyses.

We identify candidate AGN hosts from source catalogs of the 1~Ms
CDF-S and 2~Ms CDF-N X-ray images reduced and source-extracted in a
consistent fashion (Alexander et al.~2003).  These two X-ray surveys
provide the deepest views of the universe in the 0.5--8.0~keV band.
Within the respective GOODS-N(S) areas, the CDF-N(S) contains
324(223) X-ray sources down to comparable sensitivity limits of
$\approx\!1.0(1.3)\times10^{-16}$~\ergscm\ at
0.5--2.0~keV and $\approx\!7.2(8.9)\times10^{-16}$~\ergscm\ at
2--8~keV.  Coordinate-matching to the $\zband$ catalog
yields unambiguous counterparts for $>\!80$\% of the CDF sources
(F.~E.~Bauer et al.~2005, in preparation).  Many are comparatively nearby and optically
bright starbursts and ``quiescent'' galaxies contaminating our desired
X-ray--selected AGN sample.  The extensive redshift coverage of
$\zband\lesssim24$ CDF counterparts allows us to exclude these non-AGNs
with a luminosity threshold of $L(2$-$8~{\rm
keV})>10^{42}$~ergs s${}^{-1}$.  The resulting $L_X$-limited AGN sample
of 322 galaxies contains few CDF sources at $z<0.4$, so we choose this
as our lower limit for redshift-evolution analyses
(\S\ref{discsec}).

We investigate evolutionary trends in morphology and environment among
the GOODS AGN and non-AGN populations by constructing complete
volume-limited subsamples.  We estimate absolute magnitudes for the
GOODS-S galaxies by using the photometric redshift database of
Mobasher et al.~(2004), who claim an accuracy of $\Delta z/(1+z)
\lesssim 0.1$ at $\zband \lesssim 24.5$ for AGN and non-AGN alike.
Although we lack a comparable photometric redshift database for the
GOODS-N field, redshift measurements of the CDF-N sources
(Fernandez-Soto, Lanzetta, \& Yahil~1999; Cohen et al.~2000; Barger et
al.~2003) are largely complete to a similar depth ($\zband \lesssim
24.5$).  Hence, we include another 135 CDF-N sources with measured
redshifts for our volume-limited AGN sample.

To probe the evolution of the {\em optical} morphology of AGN hosts
versus the field, we compute rest-frame $B$-band (hereafter $B_0$)
quantities out to the limit $z<1.3$ accessible to our reddest filter
($\zband$).  To estimate $M_B$ and the $B_0$ structural parameters, we
linearly interpolate between $V$ and $i$ for galaxies at
$0.31<z\leq0.73$, between $i$ and $\zband$ for $0.73<z\leq1.09$, and
use $\zband$ quantities for $1.09<z<1.3$.  The redshift survey limit
of $\zband\approx24.5$ corresponds to $M_B\approx-19.5$ at $z=1.3$.
We adopt this as our limiting absolute magnitude, satisfied by 1090
GOODS-S galaxies in the non-AGN sample and another 37(42) AGN hosts
from CDF-N(S).  These volume-limited samples probe to
$\approx\!1$~mag fainter than $L_*$ and thus are not restricted to the
highest luminosity galaxies.  Coincidently, $z\sim1.3$ is the limit
for \textsl{Chandra} detection of $L(2$-$8~{\rm keV})>10^{42}$~ergs s${}^{-1}$
sources throughout the GOODS regions.  As a result, our AGN sample is
essentially complete.

\section{Concentration and Asymmetry Indices}\label{structsec}
We quantify the GOODS galaxy morphologies via non-parametric indices
of concentration $C$ and asymmetry $A$ (Conselice 2003 and references
therein).  Index $C$ scales with the ratio of radii containing 80\% and 20\%
of a source's total flux, $C\equiv 5 \log(r_{0.8}/r_{0.2})$, and
increases toward bulge-dominated morphologies.  Index $A$ is the
flux-normalized residual of the source pixels $S$ differenced with
their $180\arcdeg$-rotated counterpart $S_{180}$: $A \equiv
\min\left({{\sum_{\rm pix}{|S - S_{180}|\Big/\sum_{\rm
pix}{|S|}}}}\right) - A_0$, where $A_0$ is a background term and the
minimization is over a 0.2 pixel grid of possible centers of rotation.
While $A$ moderately increases toward disk-dominated morphologies, it is
driven to large values ($A\gtrsim0.35$) for galaxies with recent
or ongoing interaction.

Figure \ref{magconcfig} shows the $\zband$ indices $C$ (\textit{top
panel}) and $A$ (\textit{bottom panel}) for resolved sources in both GOODS
fields versus $\zband$ magnitude.  
\begin{inlinefigure}
\begin{center}
\resizebox{\textwidth}{!}
{\includegraphics{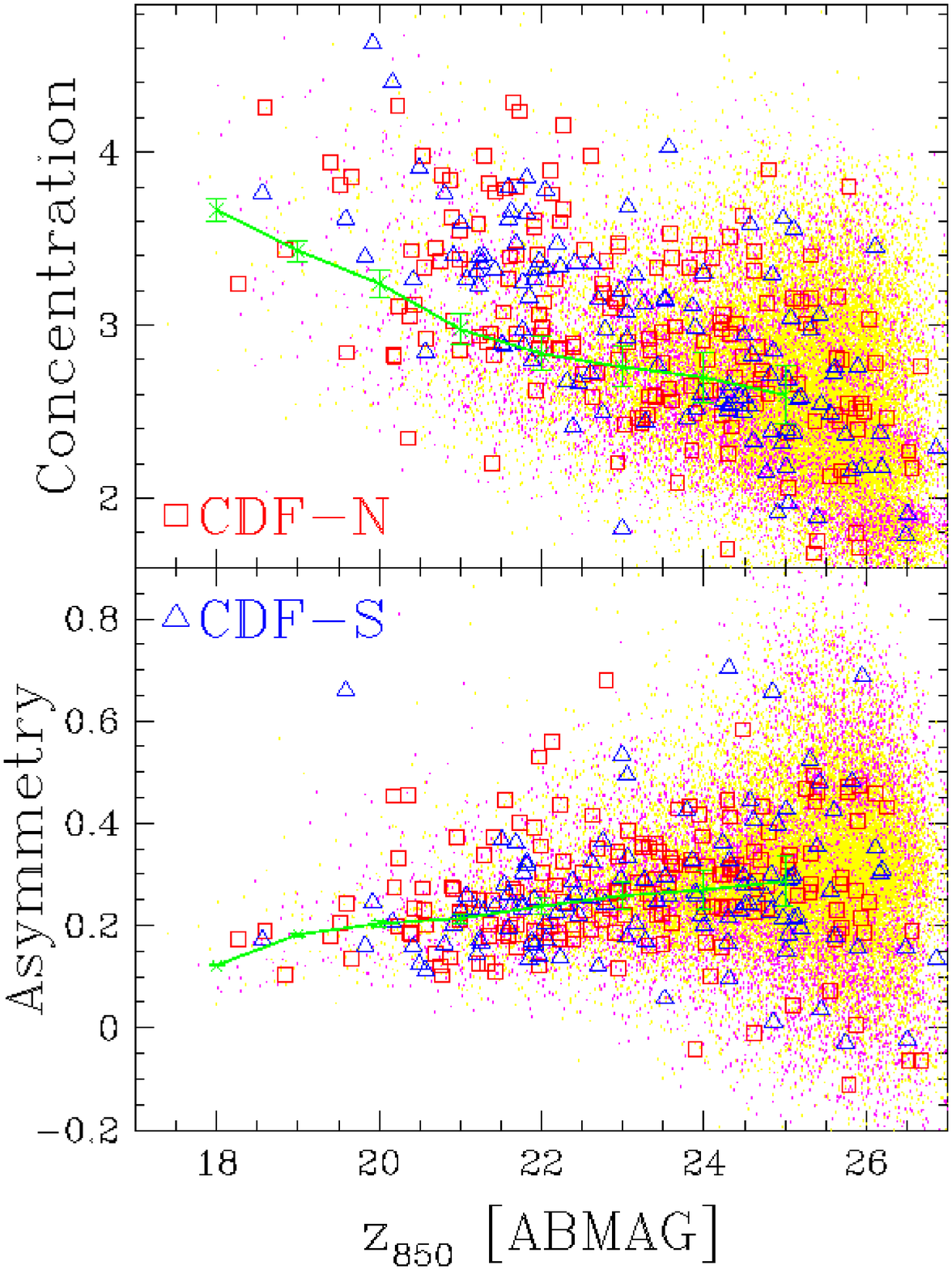}} 
\figcaption{
Concentration index (\textit{top}) and asymmetry index
(\textit{bottom}) vs.~magnitude as measured in $\zband$ for resolved
GOODS sources.  The large symbols represent the AGN sample in the
north (\textit{red squares}) and south (\textit{blue triangles}); the
small dots represent the non-AGNs in the north (\textit{yellow}) and
south (\textit{magenta}).  The connected green crosses and their error
bars denote the median values and measurement errors for the non-AGNs
in successive 1 mag bins.
\label{magconcfig}}
\end{center}
\end{inlinefigure}

\noindent
The asymmetries of the
AGN hosts (\textit{large symbols}) and non-AGN (\textit{small dots}) are not clearly
separated.  
However the AGN host concentrations clearly populate the
upper end of the field distribution throughout the regime of good
signal-to-noise ratio ($\zband \lesssim 24$).  Such large $C$-values are
preferentially associated with massive early-type galaxies, although we
note that the AGN hosts span a broad range of morphology (e.g.,
Koekemoer et al.~2002).  
The declining $C$ with host magnitude is
interpreted as an evolutionary effect; galaxies at higher redshift
(and thus generally fainter) are intrinsically less concentrated
(Conselice et al.~2003).  When comparing the $C$ distributions for
$\zband$-limited samples of AGN and non-AGN (\S\ref{concdiscsec}), we
compensate for this magnitude-dependent bias in $C$ by resampling the
non-AGN population to match the AGN hosts' magnitude distribution.

\section{Near-neighbor Frequency}\label{nearfreqsec}
In assessing the role of environment in AGN activity, we complement
the $A$ analysis with a comparison of the near-neighbor counts around
$\zband<24$ AGN hosts versus the non-AGN.  Qualifying neighbors must
satisfy both proximity and relative brightness criteria.  We
investigate two different definitions of proximity threshold $d$: one
that scales with the Petrosian radius $\rp$ of the primary galaxy,
$d<3\rp$, and another that remains fixed for all galaxies,
$d<8\arcsec$.  The latter choice, consistent with the analysis of G03,
corresponds to $54$~kpc at $z=0.6$ and varies by only $\pm\!25$\% over
the range $0.4<z<1.3$.

To limit contamination of the neighbor statistics by chance
superpositions of background galaxies, a neighbor is rejected if more
than 2~mag fainter than the comparison galaxy.  This relative
magnitude cutoff, in the presence of steeply increasing galaxy number
counts with magnitude, introduces a bias towards more neighbors around
fainter galaxies.  When comparing AGN and non-AGN near-neighbor
counts, we therefore resample the non-AGN to match the magnitude
distribution of the AGN hosts, analogous to the procedure used in
comparing $C$ distributions (\S\ref{structsec}).  Because the resolved
CDF optical counterparts typically show only minor flux contribution
from the active nucleus (G03), the relative faintness of qualifying
AGN neighbors is not significantly biased with respect to the non-AGN.
We discuss the similarity of the AGN and non-AGN near-neighbor
frequency histograms in \S\ref{asymdiscsec}.

\section{Discussion} \label{discsec}

\subsection{Bulge-dominated AGN Hosts at $z\sim0.4$-$1.3$} \label{concdiscsec}

The top left panel of Figure \ref{ksevofig} notes the
Kolmogorov-Smirnov (K-S) test probabilities for the null hypothesis
that the $\zband<23$ AGN host and non-AGN $C$-values could
be drawn from the same underlying distribution.  The quoted
probabilities are the median values from 1000 resamplings of the
non-AGN sample matched to the magnitude distribution of $z<23$ AGN
hosts to remove magnitude-dependent $C$ bias (see \S\ref{structsec}).
The $C$ distributions are highly inconsistent at the
6.4 $\sigma$ level ($P_{\rm K-S}=1.6\times10^{-10}$).  Moreover, both
the northern and the southern $\zband<23$ AGN host concentrations are
individually discrepant with their respective non-AGN counterparts at
$P_{\rm K-S} < 10^{-6}$.

\begin{inlinefigure}
\begin{center}
\resizebox{\textwidth}{!}{\includegraphics{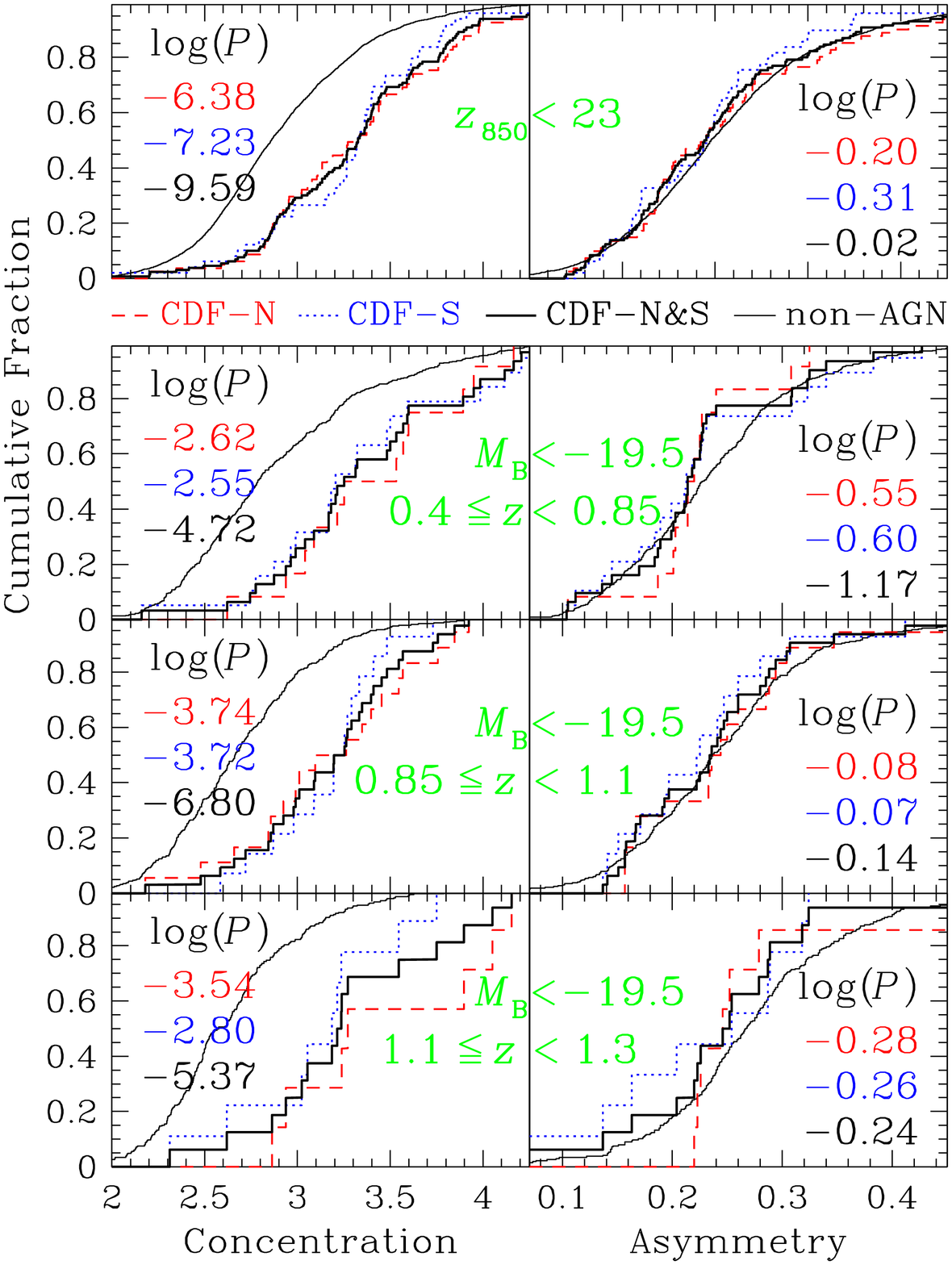}}
\figcaption{
Cumulative distribution functions of concentration index (\textit{left panels})
and asymmetry index (\textit{right panels}) for resolved GOODS AGN hosts from
the CDF-S (\textit{dotted blue line}), CDF-N (\textit{dashed red line}), and the combination of both
CDFs (\textit{thick solid black line}) for a flux-limited sample (\textit{top row}) and a
volume-limited sample divided into three redshift bins (\textit{bottom three
rows}).  The indices are compared in $\zband$ for the flux-limited
sample, and in (interpolated) rest-frame $B$ for the volume-limited
sample.  Null-hypothesis probabilities $P$ from K-S
comparisons with the control sample of non-AGN GOODS-South galaxies
(\textit{thin solid black line}) are also noted in each panel.
\label{ksevofig}}
\end{center}
\end{inlinefigure}

The median $C$ offset of $+0.5$ is consistent with the G03 measurement
based on $I<23$ sources in three \HST/WFPC2 pointings in the CDF-S
(including 25 X-ray--detected sources) and far exceeds the $\sim\!0.1$
enhancement in $C$ expected from nuclear point-source optical flux in
the AGN sample (see G03).  Moreover, $C$ enhancement by potential
circumnuclear starbursts is discounted because (1) \textit{unresolved}
starbursts reduce to the previous case for AGN point-source
contamination and (2) resolved ($\gtrsim\!500$ pc) central starbursts
with sufficient flux to bias $C$ would be inconsistent with the low
incidence of starburst-type spectral energy distributions (SEDs) among
the GOODS AGN hosts as compared to the field (Mobasher et al.~2004).

The traditional conception of ``local AGN $\equiv$ Seyfert $\equiv$
late-type host'' has been refuted by the Kauffmann et al.~(2003)
analysis of thousands of spectroscopically identified low-redshift
($z\lesssim0.3$) AGN host galaxies from the Sloan Digital Sky Survey
(SDSS).  Local AGN of all luminosities reside almost exclusively in
massive hosts with sizes, stellar mass densities, and concentration
indices similar to ordinary early-type SDSS galaxies.  The enhanced
$C$ among the GOODS AGN hosts now indicates that nuclear activity
remains preferentially associated with bulge-dominated galaxies out to
substantial lookback times ($z<1.3$).  Our $z\sim0.4$--1.3 sample
largely bridges the span of cosmic time between the quasar epoch,
where accretion-driven luminosity is dominated by high-mass SMBHs, and
the recent-epoch AGN hosts probed by SDSS.  If the locally-observed
tight correlation between SMBH mass and host $C$ (Graham et al.~2001)
similarly extends to $z\sim0.4$--1.3, then our results newly suggest
that the accretion-driven luminosity of the universe is dominated by
the most massive SMBHs at virtually all times.

The discovery of an epoch beyond which SMBH mass and host-galaxy
properties (including $C$) lose their tight correlation could strongly
constrain theories of SMBH and host-galaxy co-evolution.  To ascertain
if the $C$ discrepancy between AGN and non-AGN shows any evolution
with redshift, we divide our volume-limited sample into three redshift
bins spanning $\approx\!550$~Mpc${}^3$ each: $0.4\leq z<0.85$,
$0.85\leq z < 1.1$, and $1.1\leq z <1.3$.  The AGN host $C$ is clearly
elevated in all three bins (Fig.~\ref{ksevofig}, \textit{three bottom left
panels}), reflected in the persistently low $\log P_{\rm K-S}\sim -6.8$
to $-4.7$.  Thus our GOODS AGN host-galaxy sample populates the high
end of both the $L_X$ distribution (by construction) and the $C$
distribution throughout the range $0.4 < z < 1.3$.  Pushing this
analysis beyond $z\approx1.3$ faces multiple obstacles, including (1)
identification of obscuration-unbiased moderate-luminosity AGNs ($L_X >
10^{42}$~ergs s${}^{-1}$), requiring \textsl{Chandra} exposure depths of {\it
several} megaseconds; (2) construction of a large, complete $M_B\leq -19.5$
field sample in the so-called redshift desert; and (3)
well-resolved rest-frame optical light profiles out to meaningful
isophotes, requiring $\sim0\farcs1$ resolution $J$-band imaging at
extreme depths to overcome the sharply increasing surface brightness
dimming.
 
Although tantalizing to conclude that the Graham et al.~(2001)
SMBH-bulge correlation now persists to $z<1.3$, it is unclear whether
$L_X$ is a reasonable proxy for SMBH mass at these redshifts.  At low
redshift, AGNs with well-constrained SMBH mass are observed to have (1) $L_X
\lesssim L_{\rm Edd} \propto M_{\rm BH}$, but with a large scatter to
lower Eddington ratios (Woo \& Urry 2002) and (2) an AGN fundamental
plane in $L_X$, $L_{Radio}$, and $M_{\rm BH}$, but poor correlation
between $L_X$ and $M_{\rm BH}$ individually (Merloni et al.~2003).
However Barger et al. (2005) have recently claimed a tight
$L_X$-$M_{\rm BH}$ correlation at $z\sim1$.  Direct measurement of
the SMBH masses of GOODS galaxies would be ideal but extremely
challenging for moderate-luminosity (and often dust-obscured) AGNs at
these redshifts.

\subsection{No Merger-AGN Connection at $z\sim0.4$-$1.3$?} \label{asymdiscsec}

Unlike the disparate $C$ distributions, the asymmetry indices of the
130 resolved $\zband<23$ GOODS AGN hosts are statistically
indistinguishable from the non-AGN ($P_{\rm K-S}=0.97$).  This result
reinforces the G03 findings based on a subset of 25 X-ray--detected
sources with $I<23$.  Figure~\ref{ksevofig}
(\textit{right panels}) shows that this similarity in $A$ exists in
$B_0$ throughout all three redshift bins and separately among
northern and southern AGN subsamples.  The lowest of the K-S test
probabilities, for the combined AGN sample at $0.4\leq z<0.85$, 
does not exceed a 2 $\sigma$ rejection of the null hypothesis.

Recent/ongoing galaxy mergers in the local universe
generally have large $A$ enhancement ($A>0.35$; Conselice et
al.~2003).  Furthermore, $N$-body simulations suggest that even minor mergers
have significant $A$ enhancement up to 1~Gyr from the onset of
the merger (Walker, Mihos \& Hernquist~1996).  The lack of
differentiation between AGN and non-AGN $A$ distributions therefore
implies that recent merging/interaction is no more prevalent among the
AGNs.  This in turn argues against the hypothesis that AGN fueling is
directly linked to recent ($\lesssim\!1$~Gyr) galaxy
merging/interaction, while favoring mechanisms such as low-level gas
accretion from the intergalactic medium and/or bar instability in
disks.

\HST\ imaging of 20 nearby ($z<0.3$) quasars by Bahcall et al.~(1997)
suggested that galaxy mergers/interactions are relatively common among the
highest luminosity AGNs ($\gtrsim\!10^{44.5}$ ergs s${}^{-1}$).
However, our finding of no $A$ enhancement among more distant
($z\sim0.4$--1.3) AGNs at lower luminosities ($\lesssim\!10^{43.5}$ ergs
s${}^{-1}$) has precedent in the comparable low-redshift sample
analyzed by Corbin (2000).  The GOODS combination of deep \textsl{Chandra} and
\HST\ imaging now suggests a persistent merger-AGN disconnect
among moderate-luminosity AGNs out to
look-back times nearing the peak of AGN activity in the universe.
The limited GOODS solid angle provides too few
high-luminosity AGNs to test the Bahcall et al.~(1997)
conclusions over the same redshift range.  This may be remedied
by the wider area Galaxy Evolution from Morphology and SEDs project \HST\ imaging within the
extended CDF-S, where Sanchez et al.~(2004) have already noted a
higher incidence of merger/interaction among 15 optically selected,
$z\sim0.5$-$1.1$, type 1 AGNs with luminosities spanning the
Seyfert/quasar boundary.

Our comparison of the near-neighbor frequency histograms for AGNs and
non-AGNs yields a large $\chi^2$-test
probability of the null hypothesis both for the $d<3\rp$ threshold
[$P(\chi^2)=0.84$] and the $d<8\arcsec$ threshold [$P(\chi^2)=0.58$].
We note that the $\sim\!2\sigma$ discrepancy in near-neighbor
frequency previously observed in G03 is no longer apparent for the
substantially enlarged samples of the current work.  We conclude that
local environment, like host asymmetry, is not well correlated with
AGN activity.  This result, among moderate-redshift AGNs, now extends
similar findings of environment-AGN disconnect at low redshift from
recent analyses of the Southern Sky Redshift Survey (Maia, Machado, \&
Willmer~2003) and the SDSS (Miller et al.~2003).

\acknowledgements We acknowledge support for this work provided by
NASA through GO grants GO-09425 and GO-09583 from the Space Telescope
Science Institute, which is operated by AURA, Inc., under NASA
contract \hbox{NAS 5-26555}.  We thank the anonymous referee for 
helpful comments.

\clearpage

\end{document}